\documentclass[12pt]{iopart}
\usepackage{iopams,graphicx}

\newcommand{\be}{\begin{equation}}
\newcommand{\ee}{\end{equation}}
\newcommand{\vlk}{V_{{\rm low}\,k}}
\newcommand{\vlkh}{\overline{V}_{{\rm low}\,k}}
\newcommand{\vnn}{V_{\rm NN}}
\newcommand{\la}{\Lambda}
\newcommand{\fmi}{\, {\rm fm}^{-1}}

\begin{document}

\title{Low-momentum interactions for nuclei}

\author{Achim Schwenk}

\address{Nuclear Theory Center, 
Indiana University, 
Bloomington, IN 47408}

\begin{abstract}
We show how the renormalization group is used to construct a low-momentum 
nucleon-nucleon interaction $\vlk$, which unifies all potential 
models used in nuclear structure calculations. $\vlk$ can be directly 
applied to the nuclear shell model or to nucleonic matter without 
a $G$ matrix resummation. It is argued that $\vlk$ parameterizes 
a high-order chiral effective field theory two-nucleon force. We 
use cutoff dependence as a tool to assess the error in the truncation 
of nuclear forces to two-nucleon interactions and introduce a 
low-momentum three-nucleon force, which regulates $A=3,4$ binding 
energies. The adjusted three-nucleon interaction
is perturbative for small cutoffs.
In contrast to other precision interactions, the error due to 
missing many-body forces can be estimated, when $\vlk$ and the 
corresponding three-nucleon force are used in nuclear structure 
calculations and the cutoff is varied.
\end{abstract}



\section{Introduction}

There has been much progress over the last five years on improving
many-body methods applicable to nuclei and nucleonic matter. These
improvements are most successful in different regions of the nuclear 
chart: the Bloch-Horowitz approach for few-body systems, the 
No-Core Shell Model for light nuclei, the Coupled Cluster 
Method for the intermediate mass region, Density Functional 
Theory and effective actions for heavy nuclei, and the 
Renormalization Group approach for nucleonic matter. Although 
these microscopic many-body approaches are in principle different 
methods to diagonalize an $A$-nucleon Hamiltonian, an error 
is always introduced since it is not possible to include up 
to $A$-body forces. Thus, it is important to understand the
error of the truncation, e.g., to two-nucleon (NN), or two- and 
three-nucleon (3N) interactions, and to explore different choices 
in the nuclear force starting point.

When systems are probed at low energies, it is convenient to use
low-momentum degrees-of-freedom and replace the unresolved 
short-distance details by something simpler, without distorting 
low-energy observables. As a result, there are an infinite number of
low-energy potentials corresponding to different resolutions,
and one can use this freedom constructively to pick a convenient 
one. In nuclear physics, many issues depend on the
resolution, e.g., the strength of 3N relative to NN forces, the 
spin-orbit splitting obtained from NN only, or 
the size of exchange correlations. The change of the resolution
scale corresponds to changing the cutoff in nuclear forces, and 
thus this freedom is lost if one uses the cutoff as a fit parameter, 
or cannot vary it substantially.

In this Talk, we review results for a ``universal'' low-momentum
NN interaction, called $\vlk$. This unifies all potential models 
used in nuclear structure calculations. We present Faddeev results 
for few-body systems and show that the cutoff variation of $A=3,4$
binding energies is of the same size as results for different 
precision NN interactions, but low-momentum cutoffs are closer
to experiment. This demonstrates that cutoff independence, and 
thus model independence, in nuclear physics requires consistent 
3N forces. After augmenting $\vlk$ by chiral 3N forces, we find 
that 3N contributions are perturbative for small cutoffs. The set of 
low-momentum two- and three-nucleon interactions can be used in 
calculations of nuclear structure and reactions
and we discuss promising directions. Finally, we show that $\vlk$ and 
$G$ matrix elements are quantitatively similar, but $\vlk$ as a 
potential has a solid theoretical foundation with corresponding 
3N forces, whereas a $G$ matrix introduces uncontrolled approximations.

\section{Low-momentum nucleon-nucleon interaction}

\begin{figure}[t]
\vspace*{-1mm}
\begin{center}
\includegraphics[scale=0.4,clip=]{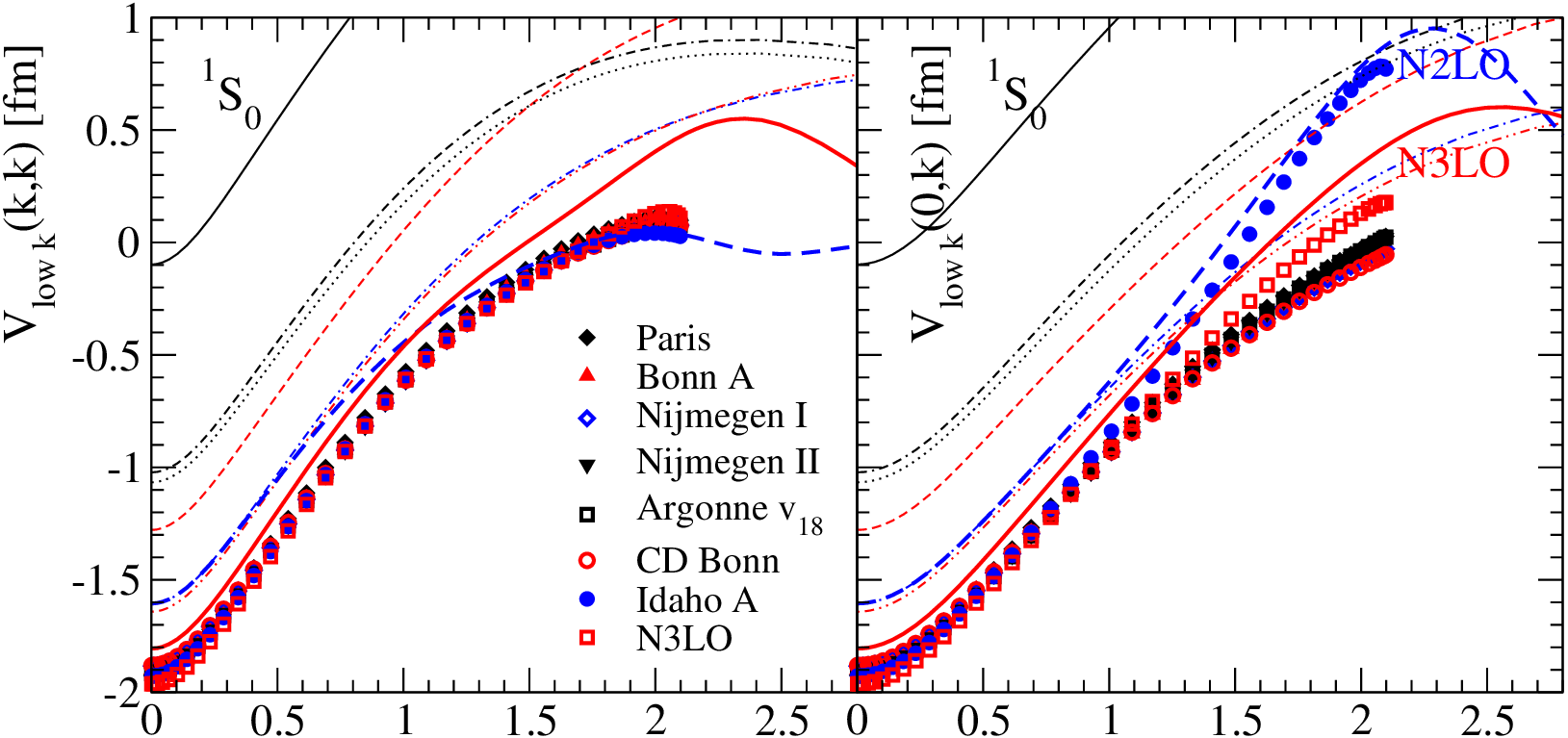}
\\[2mm]
\includegraphics[scale=0.4,clip=]{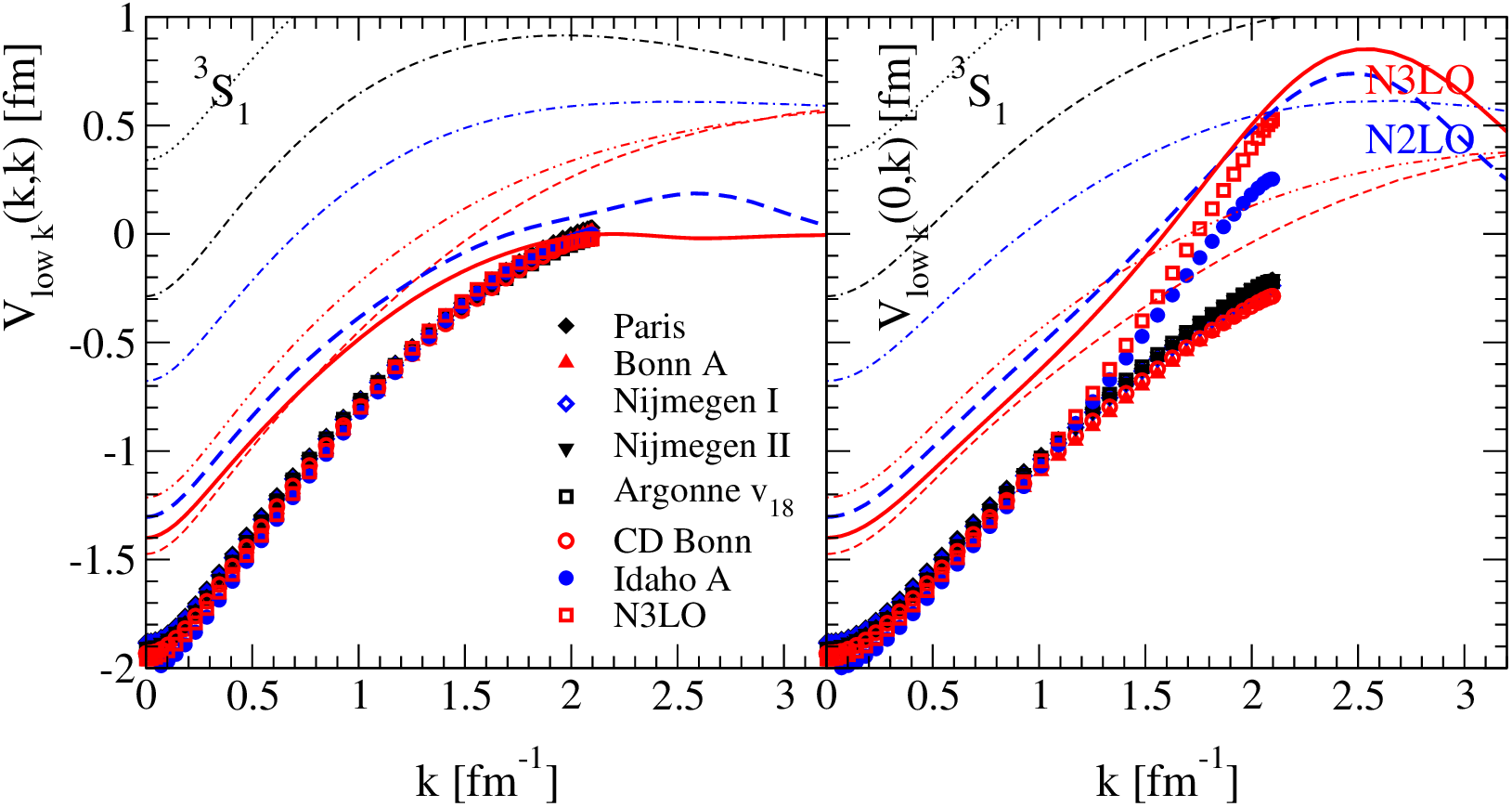}
\end{center}
\vspace*{-3mm}
\caption{\label{vlowk} Diagonal (left) and off-diagonal (right)
momentum-space matrix element for $\vlk$ (symbols) versus relative
momentum derived from different high-precision potential models
for $\la = 2.1 \fmi$. The various bare interactions are given 
as lines, with thick sold or thick dashed lines for the N2LO (Idaho A)
or N3LO interactions respectively. Results are shown for the $^1$S$_0$ 
(upper) and $^3$S$_1$ partial wave (lower figures).}
\end{figure}

Conventional precision NN interactions are well-constrained by two-nucleon 
scattering data only for laboratory energies $E_{\rm lab} \lesssim 350 \, 
{\rm MeV}$. As a consequence, details of nuclear forces are not constrained 
for relative momenta $k > 2.0 \fmi$ or distances $r < 0.5 \, {\rm fm}$.
However, all these potentials have strong high-momentum components as 
illustrated by the different lines in Fig.~\ref{vlowk}. This leads to 
model dependences and technical difficulties in many-body applications.
Starting from a given potential model $\vnn$, we have integrated out the
high-momentum modes above a cutoff $\la$ in the sense of the renormalization
group (RG)~\cite{Vlowk1,Vlowk2}. The resulting low-momentum interaction
$\vlk$ only has momentum components below the cutoff and evolves with 
$\la$ so that the low-momentum scattering amplitude $T(k',k;k^2)$ (in 
particular phase shifts and deuteron binding energy) are invariant.
Thus, in every scattering channel we have
\begin{eqnarray}
T(k',k;k^{2}) &=& \vnn(k',k) + \frac{2}{\pi} \, \mathcal{P} \int_{0}^{\infty} 
\frac{\vnn(k',p) \, T(p,k;k^{2})}{k^{2}-p^{2}} \, p^{2} dp , \\[1mm]
T(k',k;k^{2}) &=& \vlk^\la(k',k) +  \frac{2}{\pi} \, \mathcal{P} 
\int_{0}^{\la} \frac{\vlk^\la(k',p) \, T(p,k;k^{2})}{k^{2}-p^{2}} \, 
p^{2} dp .
\end{eqnarray}
In order to reproduce the low-momentum $T$ matrix for a given cutoff, 
$\vlk$ is renormalized for scattering to intermediate states with
$p>\la$. This is achieved be resumming high-momentum ladders in an 
energy-dependent effective interaction, which is the solution to
the two-body Bloch-Horowitz equation in momentum space with projector
$Q=\theta(p-\la)$. The energy dependence can then be recast as
momentum dependence by using the equations of motion. Both steps
are equivalent to the basis transformation of Lee-Suzuki.\footnote{Note
that the RG approach differs from Lee-Suzuki, as we set the $Q$ space
block of the effective Hamiltonian (which includes all model dependences)
to zero.} 
We note that
the largest effect of the renormalization is due to the first
step of integrating out the high-momentum modes (for laboratory energies 
$E_{\rm lab} \lesssim 150 \, {\rm MeV}$ the zero-energy Bloch-Horowitz
potential describes the phase shifts accurately). Both steps are
equivalent to integrating an RG equation~\cite{Vlowk1}
\be
\frac{d}{d \la} \vlk^\la(k',k) = \frac{2}{\pi} \frac{\vlk^\la(k',\la) \,
T^\la(\la,k;\la^{2})}{1-(k / \la)^{2}} .
\ee
For every cutoff $\vlk$ defines a new NN potential and a new low-momentum 
Hamiltonian
\be
H^\la_{{\rm low}\,k} = T + \vlkh^\la ,
\ee
where the cutoff acts only on the interaction and $\vlkh$ denotes a
(Okubo-) Hermitized $\vlk$ (from now on all $\vlk$ results are for 
the Hermitian $\vlkh$ and we drop the over-line). In many-body
applications, $H^\la_{{\rm low}\,k}$ will lead to different results 
from $T + \vnn$ (since unresolved interactions between any high-momentum 
nucleons are excluded).

\begin{figure}[t]
\vspace*{-1mm}
\begin{center}
\includegraphics[scale=0.45,clip=]{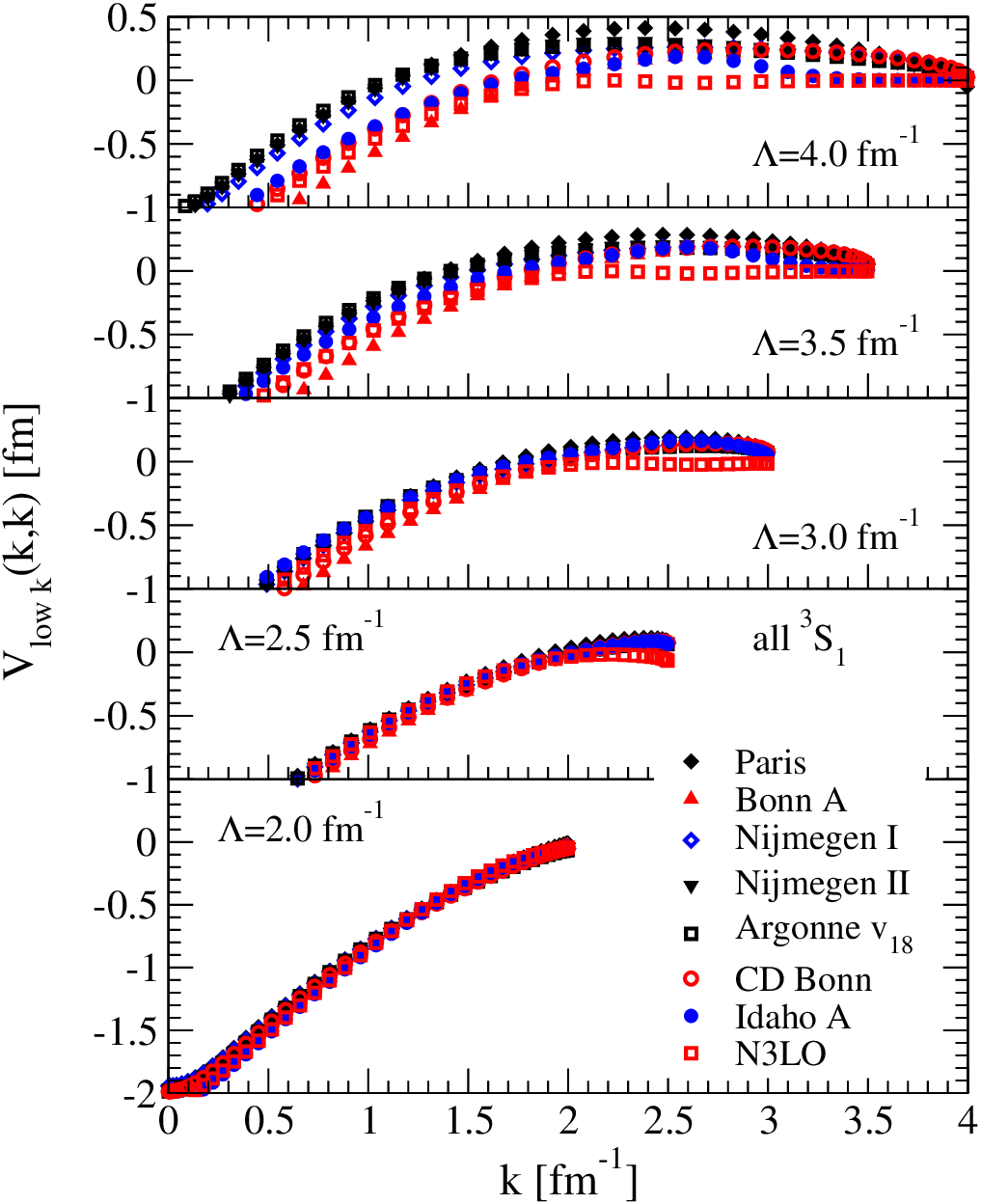}
\end{center}
\vspace*{-3mm}
\caption{\label{3s1collapse} Evolution of the diagonal $\vlk$ matrix
elements obtained from different potentials for cutoffs $\la = 
2.0 \ldots 4.0 \fmi$ versus relative momentum in the $^3$S$_1$ channel.}
\end{figure}

Our main results are shown in  Fig.~\ref{vlowk}. By performing 
an RG decimation to $\la \lesssim 2.1 \fmi$, we find that all 
NN potentials that fit the scattering data and include the 
same long-distance pion physics lead to a ``universal'' low-momentum 
interaction $\vlk$~\cite{Vlowk1,Vlowk2}. This holds for all channels
and low-momentum off-diagonal matrix elements. Note also that in
the $^1$S$_0$ channel, where there is a significant change of $\vnn$
from N2LO to N3LO, the $\vlk$ moves towards the ``universal'' curve from
N2LO to N3LO. Finally, we illustrate the collapse in 
Fig.~\ref{3s1collapse}, where we show the evolution of diagonal
$\vlk$ matrix elements from $\la = 2.0 \ldots 4.0 \fmi$. Further
results and details can be found in~\cite{Vlowk2}.

We emphasize that the renormalization of high-momentum modes is
theoretically and in practice easier in free space, before 
going to a many-body system.\footnote{We also note that the RG evolution 
is very useful for chiral effective field theory (EFT) interactions.
This is because for lower cutoffs, the phase space for intermediate 
states is smaller in nuclear structure applications. We can start 
from a chiral EFT interaction with cutoff range, e.g., $\la_\chi \sim 500 
- 700 \, {\rm MeV}$ to include the maximum known long-distance physics,
and then run the cutoff down lower. Observables are preserved under
the RG and higher-order operators are induced automatically, which is
more accurate and faster than fitting a chiral EFT truncation at the
lower cutoff.} $\vlk$ does not require a $G$ matrix resummation, which 
was introduced because of (model-dependent) high-momentum modes in 
nuclear forces. Finally, $\vlk$ is energy-independent and the cutoff 
is not a parameter (no ``magic'' value). As we demonstrate
in the next Section, the cutoff can be used to assess the error of 
a Hamiltonian truncated to two-body forces.

\section{Cutoff dependence as a tool to assess missing many-body forces}

All NN interactions have a cutoff (``P-space of QCD'') and therefore
have corresponding three- and higher-body forces. Consequently, if 
one omits the many-body forces, 3N, 4N,... observables will be 
cutoff-dependent. Using $\vlk$ all low-energy NN observables are 
cutoff-independent, and therefore, we can assess the effects of the 
omitted 3N, 4N,... forces by varying the cutoff in many-body calculations.
\begin{figure}[t]
\vspace*{-2mm}
\begin{center}
\includegraphics[scale=0.3,clip=]{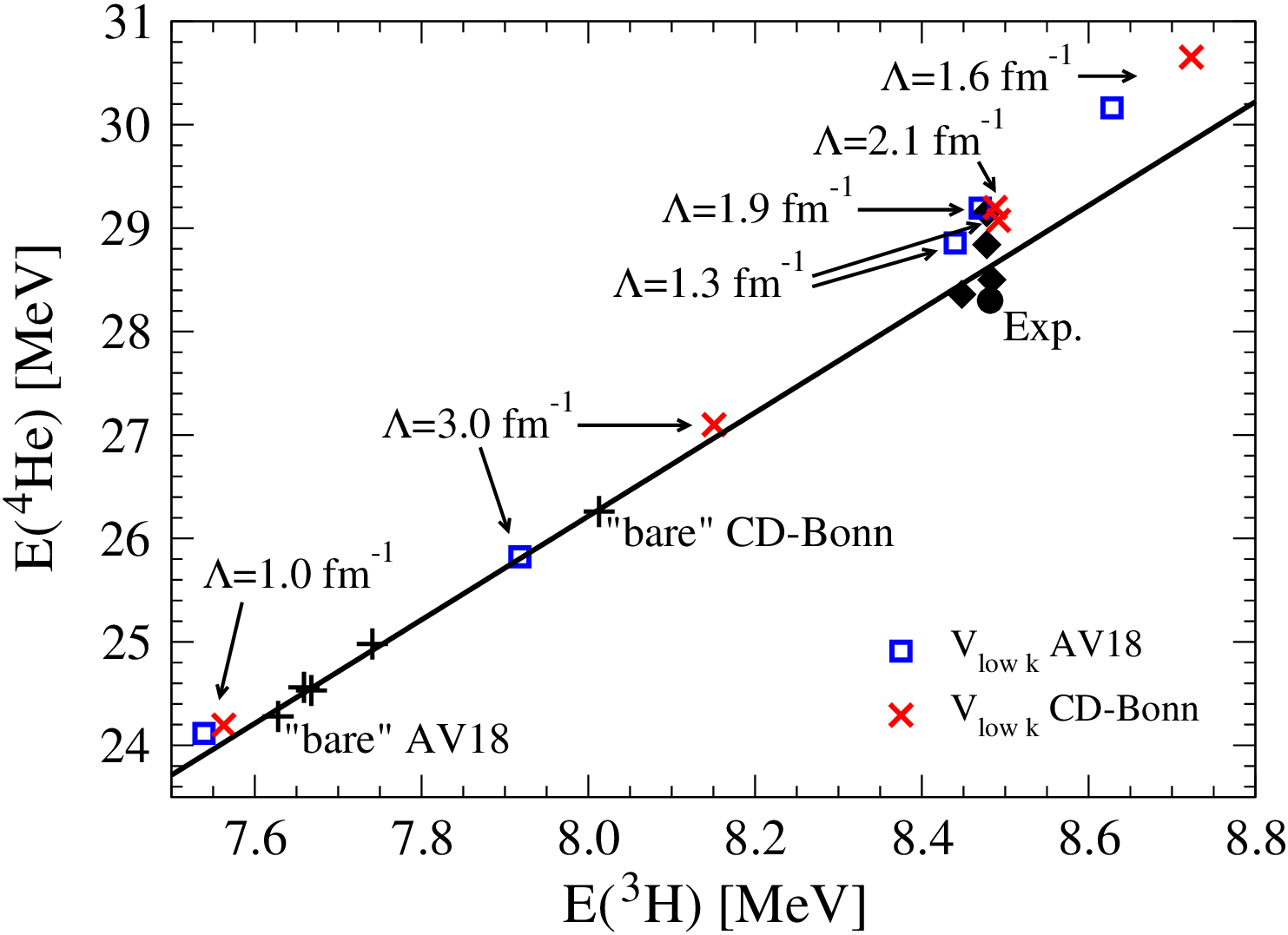}
\hspace*{2mm}
\includegraphics[scale=0.3,clip=]{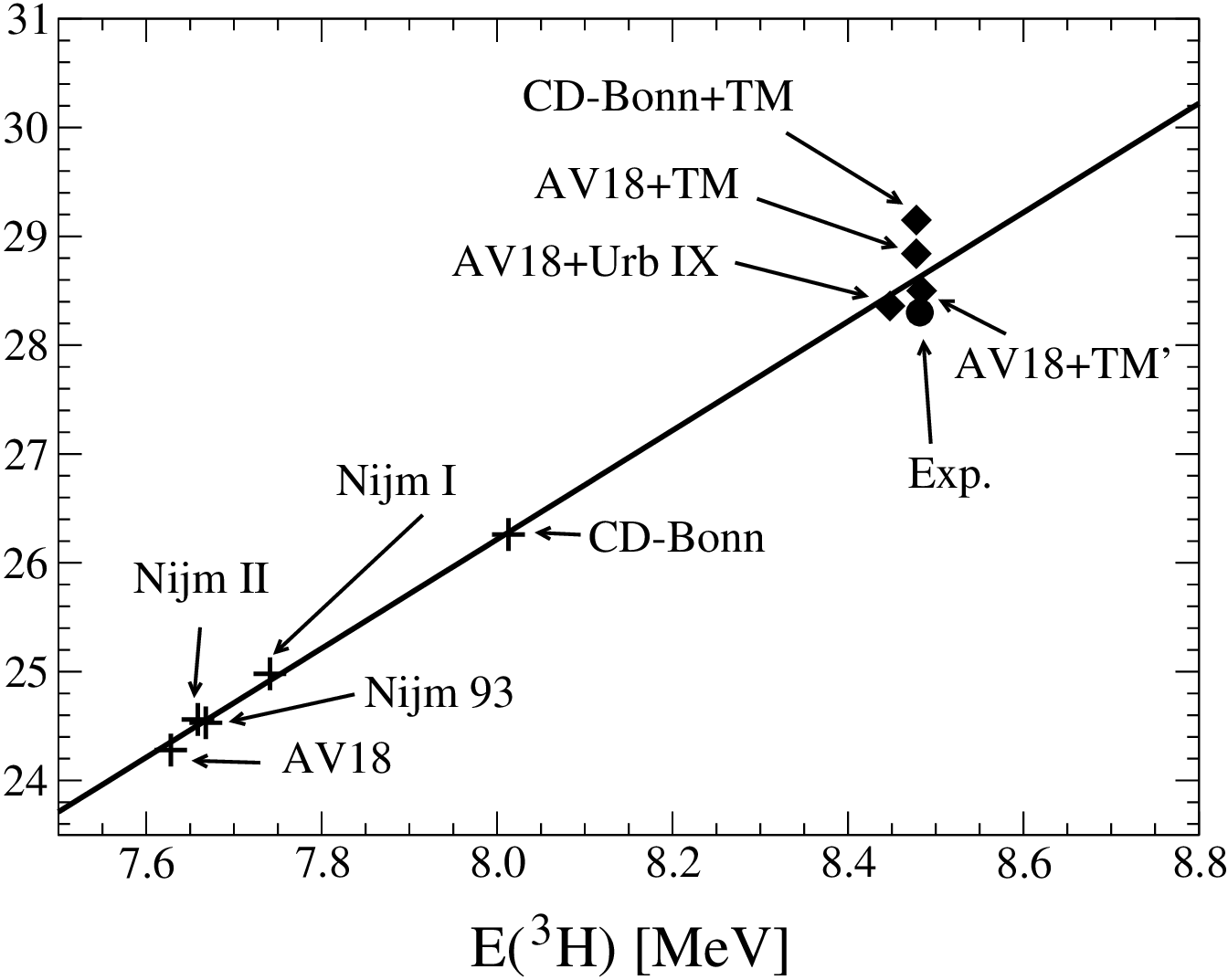}
\end{center}
\vspace*{-4mm}
\caption{\label{Tjon} Correlation of the triton and alpha particle 
binding energies. We contrast the $\vlk$ results (left figure) to
results for several modern potential models (right figure, taken 
from~\cite{Nogga}, where plusses denote NN only and diamonds NN 
with adjusted 3N forces). Results are given for the $\vlk$ derived 
from the Argonne $v_{18}$ or the CD Bonn potential. The Tjon line 
is shown as a linear fit to the NN-only results. The thick solid lines
in the left figure are interpolations from the Argonne $v_{18}$/CD
Bonn results to the respective $\vlk$ for the largest cutoff studied.}
\end{figure}
In Fig.~\ref{Tjon}, we present results for the $^3$H and $^4$He binding 
energies for a wide range of cutoffs~\cite{Vlowk3NF}. We find that the 
cutoff variations for all $\la \geqslant 1.0 \fmi$ are approximately $1 \,
{\rm MeV}$  and $5 \, {\rm MeV}$ for the $A=3$ and $A=4$ binding
energies respectively. The variation should be compared to the potential 
energy, which is $\langle \vlk^\la \rangle > 30 \, {\rm MeV}$ and $> 60 
\, {\rm MeV}$ in the two systems. The variation gives an estimate of
the contribution from many-body forces and shows that, while retaining
a $\chi^2/{\rm datum} \approx 1$, the truncation to a NN potential
alone will have associated errors of $1 \, {\rm MeV}$ and $5 \, 
{\rm MeV}$ for $^3$H and $^4$He. We also find that the cutoff dependence
predicts the linear correlation between binding energies known as the 
Tjon line. Results for reasonable 
low-momentum cutoffs $\la \sim 2.0 \fmi$ are closer 
to experiment and we observe a slight breaking off the Tjon line. Our
results demonstrate that 3N forces are inevitable for renormalization
and are similar to results obtained in the pionless EFT (see proceedings
by H.-W. Hammer). Two reference results for the $\vlk$ obtained
from the Argonne $v_{18}$ potential and $\la=1.9 \fmi$ are
$E(^3{\rm H}) = 8.47 \, {\rm MeV}$ and $E(^4{\rm He}) = 29.19 \, 
{\rm MeV}$ (for the $\vlk$ derived from the CD Bonn potential and
$\la=2.1 \fmi$, we have $E(^3{\rm H}) = 8.49 \, {\rm MeV}$ and 
$E(^4{\rm He}) = 29.20 \, {\rm MeV}$).

In the right part of Fig.~\ref{Tjon}, we also give the
potential model dependence in $A=3,4$ systems. This model dependence
is due to the probing of the unconstrained high-momentum modes. The
high-momentum modes induce three-body correlations, which are
of short-range. Therefore, at low energies these effects are inseparable 
from the effects due to omitted 3N interactions. Finally, we note
that Fujii {\it et al.} have reported similar results for $\vlk$~\cite{Fujii}, 
but conclude that large cutoffs should be used. However, this conclusion
is misguided, because it relies on reproducing the few-body binding energies
of a specific $\vnn$-only model with $\vlk$ (both without the corresponding 
3N forces). Since for every cutoff, $\vlk^\la$ is a new potential, the
few-body binding energies will be different, as expected from the right 
part of Fig.~\ref{Tjon}.

\section{Perturbative low-momentum three-nucleon interaction}

It would be extremely nice to use the RG and calculate a low-momentum 
NN and 3N interaction from different potential models. This however
has two problems. First, it requires accurate calculations of high-energy
3N scattering wave functions. Second, and more severe, with the exception
of chiral EFT interactions there is no consistency between the NN
potential models and their fitted 3N forces. As a consequence, we have
decided to adjust the leading-order chiral 3N forces to $\vlk$
for various cutoffs.
\begin{table}[t]
\resizebox{6.25in}{!}{
\begin{tabular}{l|rrrrr|rrrrr|c|c}
& \multicolumn{5}{c|}{$^3$H} & \multicolumn{5}{c|}{$^4$He} & $\max$ & $^4$He \\
\multicolumn{1}{c|}{$\la$} &
\multicolumn{1}{c}{$T$} & \multicolumn{1}{c}{$\vlk$} &
\multicolumn{1}{c}{$c$-terms} & \multicolumn{1}{c}{$D$-term} &
\multicolumn{1}{c|}{$E$-term} & \multicolumn{1}{c}{$T$} &
\multicolumn{1}{c}{$\vlk$} & \multicolumn{1}{c}{$c$-terms} &
\multicolumn{1}{c}{$D$-term} & \multicolumn{1}{c|}{$E$-term} &
\multicolumn{1}{c|}{$|V_{\rm 3N}/\vlk|$} & 
\multicolumn{1}{c}{$k_{\rm rms}$} \\ \hline
$1.0$ & $21.06$ & $-28.62$ & $0.02$ & $0.11$ & $-1.06$ & 
$38.11$ & $-62.18$ & $0.10$ & $0.54$ & $-4.87$ & $0.08$ & $0.55$ \\
$1.3$ & $25.71$ & $-34.14$ & $0.01$ & $1.39$ & $-1.46$ & 
$50.14$ & $-78.86$ & $0.19$ & $8.08$ & $-7.83$ & $0.10$ & $0.63$ \\
$1.6$ & $28.45$ & $-37.04$ & $-0.11$ & $0.55$ & $-0.32$ & 
$57.01$ & $-86.82$ & $-0.14$ & $3.61$ & $-1.94$ & $0.04$ & $0.67$ \\
$1.9$ & $30.25$ & $-38.66$ & $-0.48$ & $-0.50$ & $0.90$ &
$60.84$ & $-89.50$ & $-1.83$ & $-3.48$ & $5.68$ & $0.06$ & $0.70$ \\ 
$2.5(a)$ & $33.30$ & $-40.94$ & $-2.22$ & $-0.11$ & $1.49$ &
$67.56$ & $-90.97$ & $-11.06$ & $-0.41$ & $6.62$ & $0.12$ & $0.74$ \\
$2.5(b)$ & $33.51$ & $-41.29$ & $-2.26$ & $-1.42$ & $2.97$ &  
$68.03$ & $-92.86$ & $-11.22$ & $-8.67$ & $16.45$ & $0.18$ & $0.74$ \\
$3.0(*)$ & $36.98$ & $-43.91$ & $-4.49$ & $-0.73$ & $3.67$ & 
$78.77$ & $-99.03$ & $-22.82$ & $-2.63$ & $16.95$ & $0.23$ & $0.80$
\end{tabular}}
\caption{\label{3Nparts} Expectation values of the kinetic energy ($T$), 
$\vlk$ and the different 3N contributions (long-range $2 \pi$-exchange
($c$-terms), $1 \pi$-exchange part ($D$-term) and contact interaction
($E$-term)) for $^3$H and $^4$He. All energies are in MeV and momenta 
are in fm$^{-1}$. $(a)$ and $(b)$ denote two possible solutions for 
$\la = 2.5 \, {\rm fm}^{-1}$ and $(*)$ indicates that the $^4$He fit is 
approximate, for details see~\cite{Vlowk3NF}. 
In addition, we give the ratio of maximum 3N to $\vlk$ contribution and 
an average relative momentum $k_{\rm rms}$.}
\end{table}
The motivation for this is that at low energies, all phenomenological 3N
forces due to meson exchanges and high-momentum N, $\Delta$,... intermediate
states collapse to this operator form. Therefore, it is reasonable to
use the operator form constrained by chiral EFT and adjust the coupling
constants to $\vlk$. Moreover, cutoffs in $\vlk$ and chiral potentials 
are very similar. Both are low-momentum interactions, where only 
pion exchanges are explicitly resolved. When we start from chiral 
interactions and run the cutoff down lower, we find that they also collapse 
to the ``universal'' $\vlk$, which indicates that $\vlk$ parameterizes
a higher-order EFT interaction with sharp-cutoff regularization.

In chiral EFT, the leading-order 3N interaction enters at N2LO and
consists of a long-range $2 \pi$-exchange, an intermediate-range
$1 \pi$-exchange part and a short-range contact 
interaction~\cite{chiral3NF1,chiral3NF2}. There are five coupling constants: 
three low-energy $c$ constants in the 
$2\pi$ part, as well as $D$- and $E$-term couplings in the $1 \pi$
and contact term respectively. A possible determination of the 
$c$ constants is through a NN partial wave analysis, which includes
the long-distance $1 \pi$ and $2 \pi$ physics in the interaction.
This has been carried out by the Nijmegen group, and we take their
values for the $c$ constants~\cite{const}, which is most in keeping 
with our results that $\vlk$ is strongly constrained by the scattering 
data. Since the $c$ constants parameterize low-energy $\pi N$ physics,
their values are independent of the cutoff used to regularize nuclear
forces.

We then adjust the two $D$- and $E$-term couplings to the $^3$H and
$^4$He binding energies for different cutoffs (We note that it may
be better to adjust the 3N force to $^3$H and a heavier system, say
$^{16}$O, when the latter can be calculated more accurately in the 
future). For cutoffs $\la
\lesssim 2.0 \fmi$ we find linear dependences in the fitting,
which are consistent with a perturbative $D$- and $E$-term
contribution $E(^3{\rm H}) = E(\vlk + c{\rm-terms}) + c_D \langle
D{\rm-term} \rangle + c_E \langle E{\rm-term} \rangle$ ($c_D$ and
$c_E$ are the coupling constants and $\langle \ldots \rangle$
denote the matrix elements of the operators). This has been
checked explicitly and also for the $c$-terms. For $\vlk$ and
all studied cutoffs $\la \lesssim 2.0 \fmi$, we find that the adjusted
3N forces are perturbative~\cite{Vlowk3NF}, by which we 
mean $\langle \Psi^{(3)} | V_{\rm 3N} | \Psi^{(3)} \rangle \approx
\langle \Psi^{(2)} | V_{\rm 3N} | \Psi^{(2)} \rangle$, where $| \Psi^{(n)}
\rangle$ are exact solutions including up to $n$-body forces.
We use the operator form of the chiral 3N force given in Eq.~(2) 
and Eq.~(10) in~\cite{chiral3NF2}, multiplied by an exponential
regulator $\exp[-((p^2+3q^2/4)/\la^2)^4]$ with the same cutoff value
as in $\vlk$ ($p$ and $q$ are Jacobi momenta). The high power 
in the exponent yields a behavior similar to a sharp cutoff.
The two parameters of the 3N force fit to $\vlk$ are tabulated 
for a wide range of cutoffs in~\cite{Vlowk3NF}. We note that
it is non-trivial that a fit solution of the leading-order 
chiral 3N form with realistic $c$ constants exists.

Next, we present the different contributions to the triton and 
alpha particle binding energies in Table~\ref{3Nparts}. Assuming
that the kinetic energy is due to independent particle pairs,
we can use this to obtain an average relative momentum $k_{\rm rms}
= \sqrt{\langle k^2 \rangle} \approx m T /(A-1)$, when $\vlk$ is
used in these systems. Over the range of cutoffs in Table~\ref{3Nparts},
we find $k_{\rm rms} \approx 0.55 \ldots 0.80 \fmi$ for $^4$He 
($0.50 \ldots 0.67 \fmi$ for $^3$H). Although not observable, it
is reassuring that $k_{\rm rms} \ll \la$ and also intriguing that for
all low-momentum cutoffs $k_{\rm rms} \sim m_\pi$, as expected in
chiral EFT. We also find that the non-linearities in the fitting
for larger cutoffs $\la \gtrsim 2.5 \fmi$ lead to a ratio 
of the maximum 3N to $\vlk$ contribution $\approx 0.2$. In the
chiral counting, 3N contributions are on the order of $(Q/\la)^3$
relative to the NN force, where $Q$ is a typical momentum in the
system. With $Q \sim k_{\rm rms} \sim m_\pi$, we find $(Q/\la)^3 \approx
0.05$ for $\la \sim 2.0 \fmi$ and 
thus the $0.2$ ratio at larger cutoffs is beyond this
expectation. We take this as an indication that the leading-order
chiral 3N force is insufficient for larger cutoffs, where more
physics is resolved, or that one enters the non-linear regime
of a limit cycle (see proceedings by A. Nogga). We also observe
that the $c$-terms and the $E$-term increase with increasing cutoff and cancel.
This is expected since the $E$-term renormalizes all divergences
in the 3N system. 

It is important to note that the 3N contributions, while perturbative
for all cutoffs $\la \lesssim 2.0 \fmi$, increase by a factor
$\sim 5$ from $A=3$ to $A=4$. This density dependence 
leads to saturation in nuclear
matter~\cite{nuclmat}. Finally, for further details on $\vlk$ in 
few-nucleon systems and the low-momentum 3N force see~\cite{Vlowk3NF} 
and proceedings by A. Nogga.

\section{Harmonic-oscillator matrix elements and $G$ matrix comparison}

\begin{figure}[t]
\vspace*{-1mm}
\begin{center}
\includegraphics[scale=0.4,clip=]{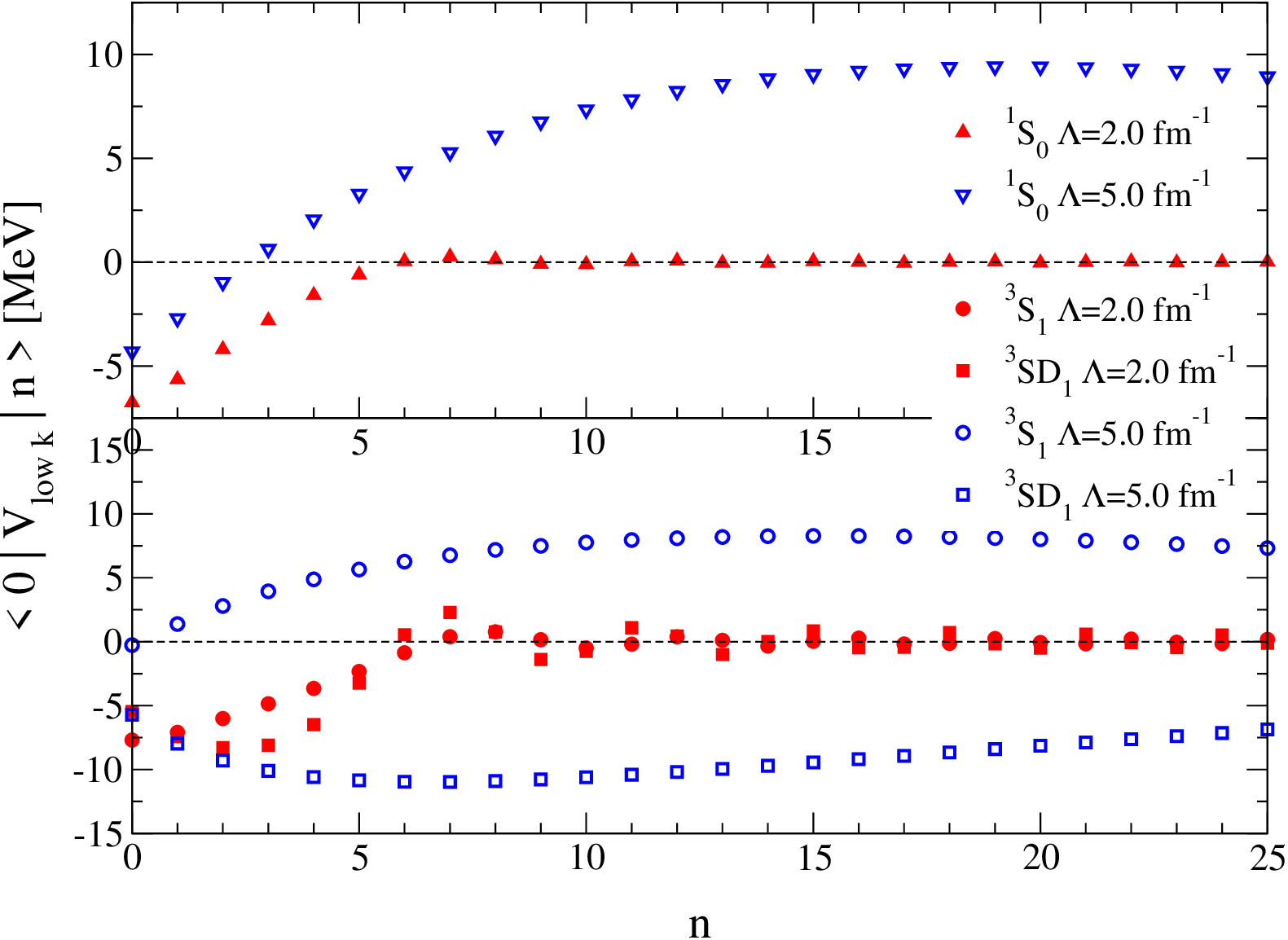}
\end{center}
\vspace*{-3mm}
\caption{\label{relho} Relative harmonic-oscillator matrix elements
$\langle n'=0 \, l' | \vlk | n \, l \, S \, J \rangle$ versus
radial quantum number $n$ for a low-momentum cutoff $\la = 2.0 
\, {\rm fm}^{-1}$ and a $1.0 \, {\rm GeV}$ 
cutoff $\la = 5.0 \, {\rm fm}^{-1}$. 
Results are shown for the S-wave matrix elements. In both cases, $\vlk$ is 
obtained from the Argonne $v_{18}$ potential and $\hbar \, \omega = 
14 \, {\rm MeV}$.}
\end{figure}

The advantage of using low-momentum interactions in many-body applications 
is that $\vlk$ is a soft interaction, without a strong core at short 
distances. Therefore, $\vlk$ does not couple strongly to high-lying states
and can be used in small model spaces. This makes it for the first time 
possible to start directly from a precision NN interaction for applications 
in nuclear structure or reactions.

The benefit of a lower cutoff in shell model applications is shown
in Fig~\ref{relho}, where we compare S-wave relative harmonic-oscillator
matrix elements for a low-momentum cutoff $\la = 2.0 \fmi$ and a
$1.0 \, {\rm GeV}$ cutoff $\la = 5.0 \fmi$. We find that for low-momentum
interactions the matrix elements decrease quite rapidly and become
small for $|n - n'| \sim 10$. This is not the case for interactions
with high-momentum components, which require basis states up to $\sim
50$ shells for convergence. Fig.~\ref{relho} clearly shows that strong 
high-momentum modes are only poorly represented in a shell model basis.

\begin{figure}[t]
\vspace*{-1mm}
\begin{center}
\includegraphics[scale=0.35,clip=]{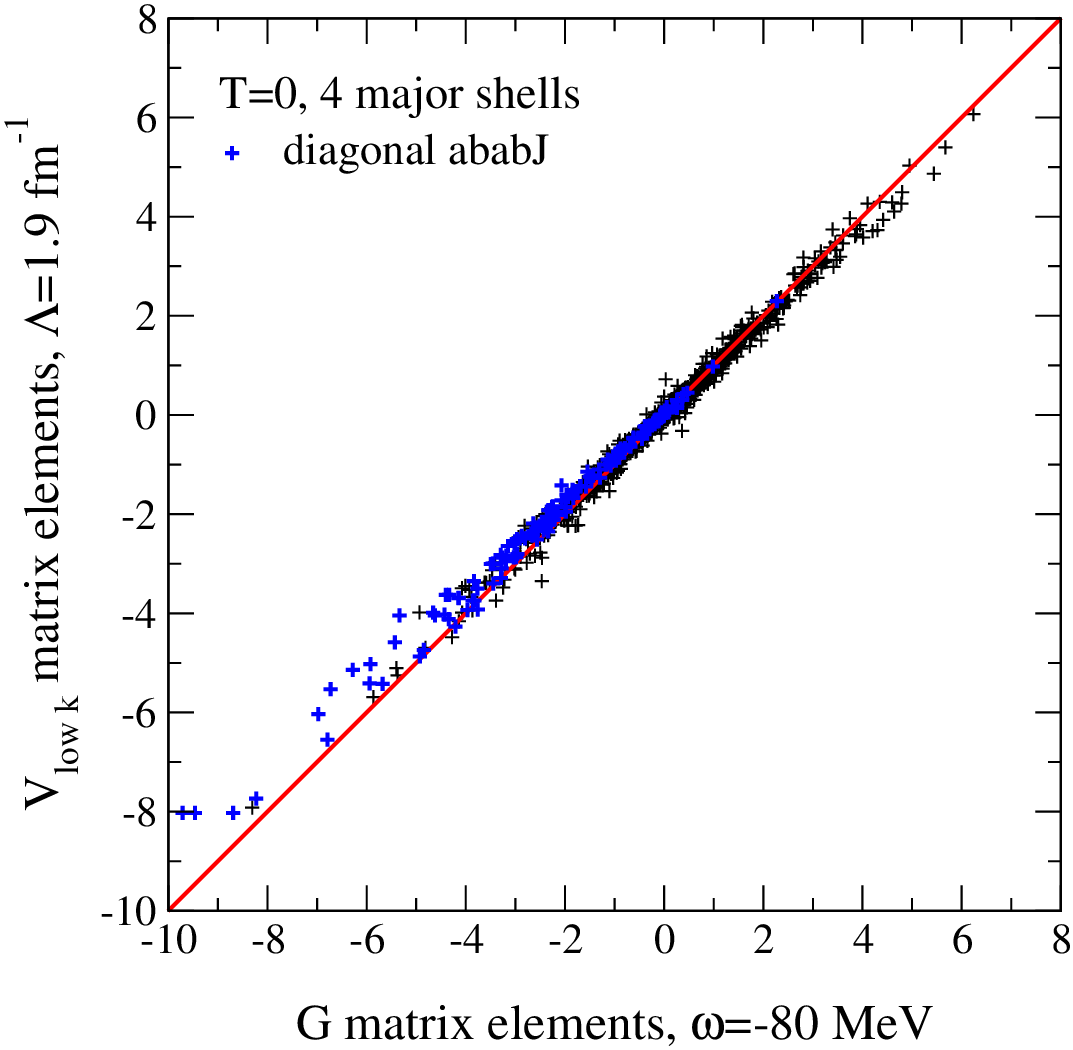}
\hspace*{2mm}
\includegraphics[scale=0.35,clip=]{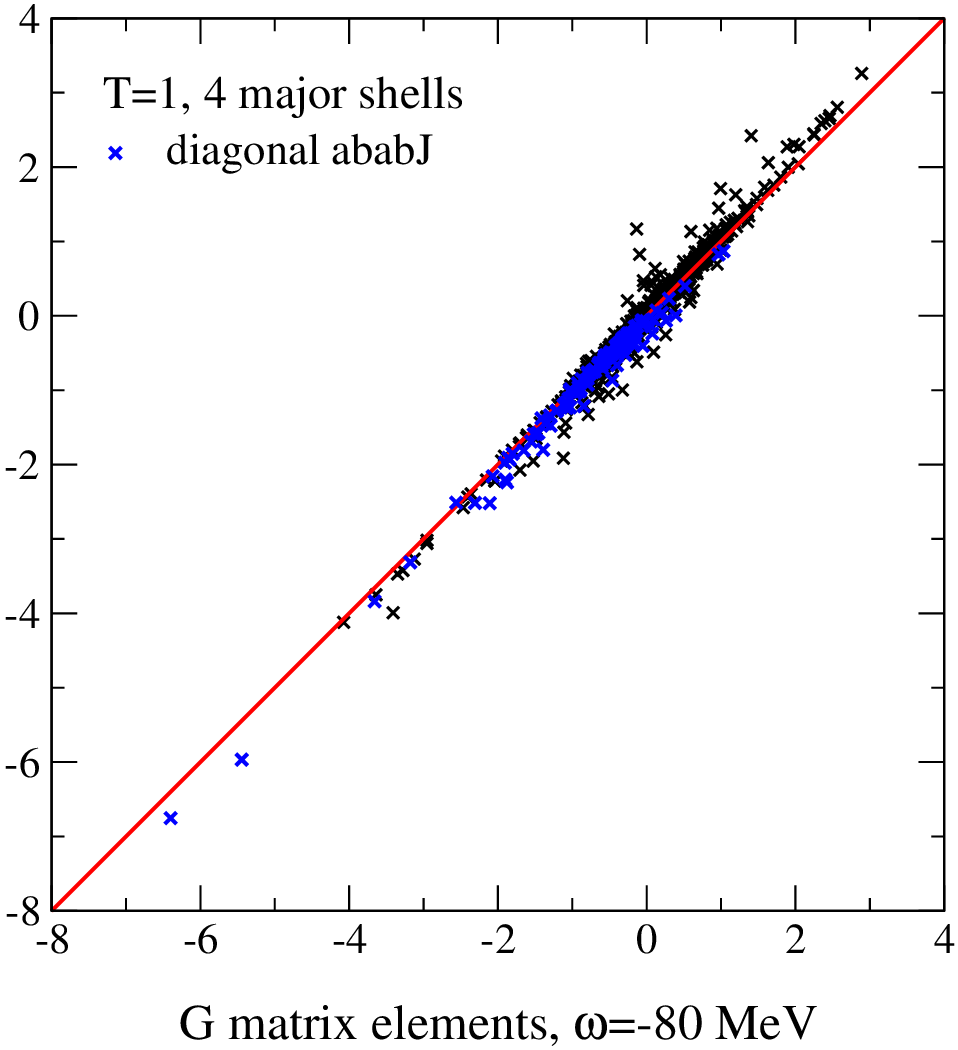}
\end{center}
\vspace*{-3mm}
\caption{\label{corrplot}Correlation plots between
$\vlk$ and $G$ matrix elements in 4 major shells. The matrix 
elements $\langle a \, b \, | \ldots | \, c \, d \, J \, T \rangle$ are in 
${\rm MeV}$ for $\hbar \, \omega = 14 \, {\rm MeV}$. We also distinguish 
the diagonal elements related to the monopole interaction and thus 
nuclear binding. $\vlk$ is derived from the Argonne $v_{18}$ potential
and the $G$ matrix is for Idaho A computed with a rectangular 
Pauli operator for 4 major shells and starting energy $- 80 \, 
{\rm MeV}$~\cite{David} (for other correlation plots see proceedings 
by B.A. Brown).}
\end{figure}

In conventional approaches to shell-model effective interactions, the 
cores are tamed by performing a ladder resummation of a $\vnn$ model
to obtain a $G$ matrix. However, the $G$ matrix resummation introduces
an uncontrolled starting-energy dependence and requires further
approximations in practice. Moreover, there is no theory for the 
starting energy, since the Bloch-Horowitz energy self-consistency 
is lost when one restricts the effective interaction to two-body
in an $A$-body system. In Fig.~\ref{corrplot}, we compare $\vlk$ to
$G$ matrix elements in four major shells.
We find that in both $T=0,1$ channels the matrix elements are very similar.
The biggest differences are on the diagonal $T=0$ matrix elements. This
comes as no surprise since the diagonal matrix elements are related
to the monopole interaction, responsible for 
most of the nuclear binding. Here, we
expect the largest effect of the low-momentum 3N force. A detailed
and more quantitative comparison will be presented in~\cite{pheno},
where we also study the phenomenology of low-momentum interactions and 
point out where calculations including 3N forces are most needed.
We note that similar correlations between $\vlk$ and a $G$ matrix
hold for different cutoffs and different $\hbar \omega$ (up to a simple 
scaling)~\cite{pheno}. Here, we only want 
to show that $\vlk$ can be used directly 
in nuclear structure applications, and emphasize that $\vlk$ has a 
well-defined theoretical basis with perturbative, low-momentum 3N forces.

\section{Summary and outlook}

In this Talk, we have reviewed advances in constructing low-momentum
interactions, which can be used directly in many-body applications.
Our results show that, for low-momentum cutoffs, nuclear forces are
well constrained and that difficult-to-handle cores are not needed
to reproduce the NN scattering data. We have shown how the RG can be
used to construct a model-independent low-momentum interaction, which
unifies all precision potential models used in nuclear structure
calculations. We believe that $\vlk$ is a very useful for nuclear
many-body problems for the following reasons:
\begin{enumerate}
\item It is possible to vary the cutoff in $\vlk$ (or $\vlk$ and
adjusted 3N interactions) over a wide range. This enables one to
estimate an error due to omitted many-body forces (or omitted higher-order 
3N, 4N,... interactions). In this way, it is possible to vary the
cutoff for interactions with singular pion exchanges and obtain
an error control in many-body calculations as in the pionless EFT.
\item We have shown that 3N forces are required by renormalization
and that adjusted low-momentum 3N interactions are perturbative 
for cutoffs $\la \lesssim 2.0 \fmi$. This should considerably
simplify including 3N forces in nuclear structure applications,
e.g., shell model interactions, coupled cluster theory or nucleonic
matter.
\item Low-momentum interactions bind nuclei 
in Hartree-Fock~\cite{pertnuclei}, in 
contrast to all other microscopic NN interactions. Consequently,
exchange correlations are smaller starting from low-momentum
forces, and a quantitative derivation of a nuclear density functional
seems possible~\cite{DFT}. Moreover, preliminary results indicate that nuclear 
matter with $\vlk$ and 3N forces is perturbative~\cite{nuclmat}, for
neutron matter the Hartree-Fock equation of state is very 
reasonable~\cite{indint}.
\item $\vlk$ as a potential provides a well-defined starting
point for microscopic calculations of effective interactions
for heavier nuclei in small model spaces.
\end{enumerate}

\noindent
Applications of $\vlk$ that were not discussed in this Talk range from
valence particle nuclei~\cite{O18} to quasiparticle interactions and 
pairing in neutron matter~\cite{RGnm,tensor}.

We close with a recommendation of how to start using $\vlk$ and
some priorities for future research. If one wants to use 
$\vlk$ as a new potential without varying the cutoff, we suggest to 
use cutoffs near $\la = 2.0 \fmi$ (for the $\vlk$ derived
from the Argonne $v_{18}$ (CD Bonn) potential the triton binding
energy is accidentally reproduced for $\la = 1.9 \fmi$ ($\la
= 2.1 \fmi$)). We would take these as first cutoff values but 
stress that the 3N force never vanishes. For future investigation,
it would be extremely promising to study the cutoff variation of 
nuclear spectra and convergence properties using $\vlk$ with 3N forces in 
the No-Core Shell Model or Coupled Cluster Theory. Varying the cutoff
will be a powerful tool to provide theoretical error estimates for
extrapolations towards the drip lines, where one cannot compare to
experiment. A chart for the spectra of light nuclei with theoretical
error bars would be wonderful. Finally, more insight on the effects
of 3N forces will come from shell model calculations using $\vlk$ 
with perturbative 3N forces, especially where a two-body $G$ matrix fails.

\vspace*{2mm}
\noindent
It is a pleasure to thank my collaborators Scott Bogner, Gerry Brown, 
Bengt Friman, Dick Furnstahl, Chuck Horowitz, Tom Kuo, Andreas Nogga,
Janos Polonyi and Andres Zuker for many discussions. 
This work is supported by the DOE under grant No. DEFG 0287ER40365 and 
the NSF under grant No. NSF--PHY 0244822.

\end{document}